\documentclass[11pt]{scrartcl}

\usepackage{amsmath}
\usepackage{upgreek}
\usepackage{graphicx}
\begin{document}

\title{\LARGE{Experiments on transformation thermodynamics:\\Molding the flow of heat}}
\author{\normalsize{Robert~Schittny$^1$, Muamer~Kadic$^1$, Sebastien~Guenneau$^2$, and Martin~Wegener$^{1,3}$}}

\date{}
\maketitle

\begin{center}
\textit{\noindent$^1$Institute of Applied Physics, Karlsruhe Institute of Technology (KIT), 76128~Karlsruhe, Germany}

\textit{\noindent$^2$Institut Fresnel, CNRS, Aix-Marseille Universit\'e, Campus Universitaire de Saint-J\'er\^ome, 13013 Marseille, France}

\textit{\noindent$^3$Institute of Nanotechnology, Karlsruhe Institute of Technology (KIT), 76128~Karlsruhe, Germany}
\end{center}

\begin{abstract}
It has recently been shown theoretically that the time-dependent heat conduction
equation is form-invariant under curvilinear coordinate transformations. Thus,
in analogy to transformation optics, fictitious transformed space can be mapped onto
(meta-)materials with spatially inhomogeneous and anisotropic heat-conductivity
tensors in the laboratory space. On this basis, we design, fabricate, and characterize a
micro-structured thermal cloak that molds the flow of heat around an object in a metal plate. This allows for transient protection of the object
from heating, while maintaining the same downstream heat flow as without object
and cloak.
\end{abstract}

Transformation optics is a design tool that enables steering the flow of electromagnetic waves \cite{Pendry2006, Leonhardt2006, Smith2006}, liquid waves \cite{Farhat2008,Fang2011}, elastodynamic waves \cite{Milton2006, Cummer2007, Norris2008, Brun2009, Stenger2012}, or quantum mechanical matter waves \cite{Zhang2008,Greenleaf2008} in an unprecedented manner. In essence, for each of these examples, curved space is mapped onto a (meta-)material distribution \cite{Pendry2006,Leonhardt2010,Milton2006}. For this mapping to be possible, the underlying equations need to be form-invariant under curvilinear coordinate transformations.

Can this basic idea also be translated to other problems such as electrical conduction, heat conduction, or particle diffusion? The answer may not appear obvious because the time-dependent parabolic differential equations underlying all of these problems are distinctly different from the hyperbolic differential equations which govern wave phenomena. For example, basic phenomena like reflection, scattering, shadows, polarization, or interference are crucial aspects in many wave systems, but are completely absent in scalar parabolic differential equations. However, conformal maps have already been used for more than a century in two-dimensional stationary potential-flow problems in hydromechanics to transform boundaries (not materials though) \cite{Lamb1879}. The case of time-independent electric conduction was studied theoretically in 1984 by Kohn and Vogelius (following an observation by Tartar) \cite{Kohn1984} and in 2003 by Greenleaf and coworkers \cite{Greenleaf2003}. The authors concluded that the two-dimensional tomography problem is not unique. This means that one cannot distinguish between, e.g., (i) a homogeneous conductive plate and (ii) a plate with a hole and a specially tailored conductivity distribution around it by measuring the electric resistance between any number of pairs of points outside the central region. Today, this “non-uniqueness” is commonly referred to as “cloaking”. Cloaking represents a benchmark example for any transformation approach. Only quite recently, experiments have been published for steady-state electric cloaking \cite{Yang2012}. Experiments have also been published \cite{Narayana2012} within the time-independent limit of the corresponding (elliptic) continuity equation for the heat current density (comprising a heat shield, but no cloak); also see previous theoretical work \cite{Lurie1988}. However, the more general thermo\textit{dynamic} problem has only been treated theoretically so far \cite{Guenneau2012}. In the dynamic case, the spatial profile of the effective specific heat enters explicitly, whereas it is strictly irrelevant in the static case. Interestingly, the time-dependent heat-conduction equation is mathematically equivalent to the time-dependent particle-diffusion equation. 

\begin{figure}
  \centering
  \includegraphics[scale=0.75]{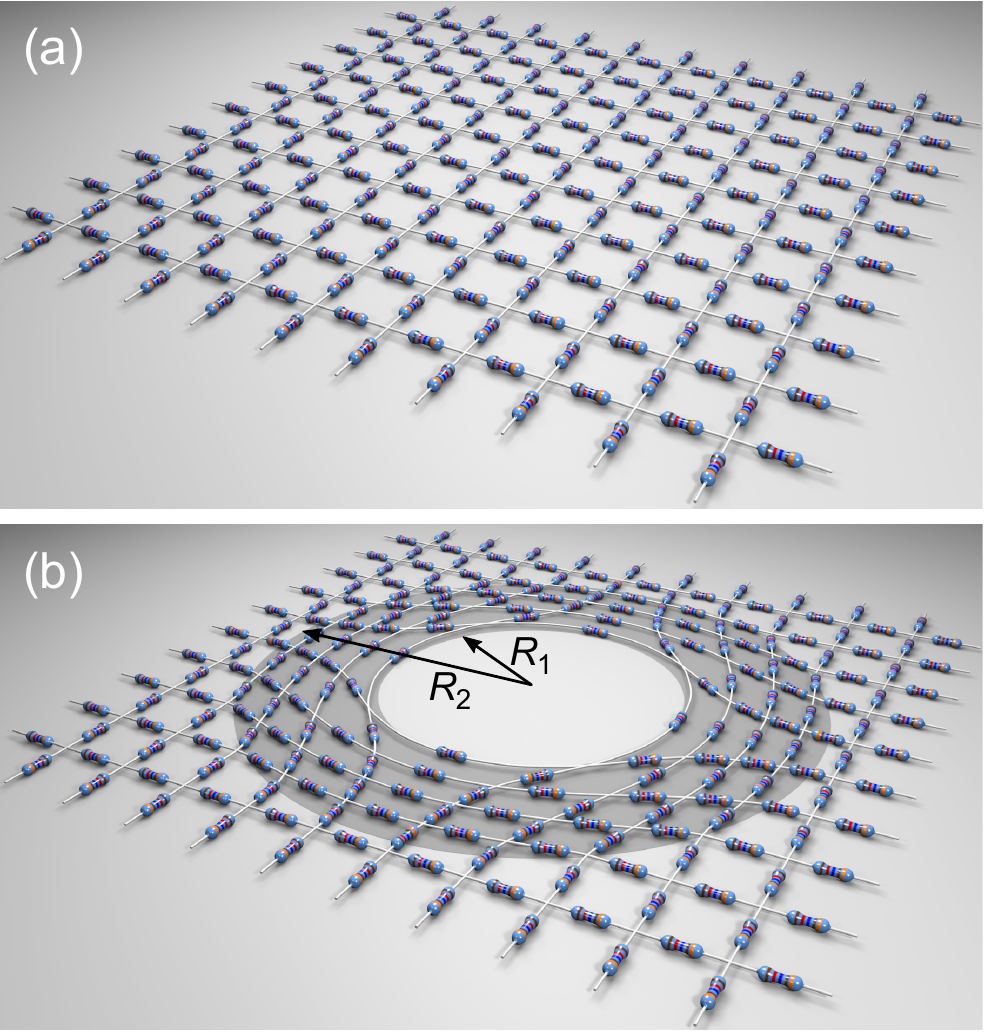}
  \caption{(a) Square lattice of electrical or thermal resistors. (b) Illustration of the curvilinear coordinate transformation underlying the cloak to be shown in Fig.~\ref{fig2_structure_design}. The cloak region is shaded in gray with the inner and outer radii $R_1$ and $R_2$ marked. Note that the effective local conductivity varies in both the azimuthal and the radial direction. For an observer measuring heat or electrical flux outside the shaded region, configurations (a) and (b) are indistinguishable. This is the essential idea of thermal cloaking.}
  \label{fig1_resistor-grid}
\end{figure}

In this paper, we experimentally realize a thermodynamic cloak for what we believe to be the first time. Movies of the transient temperature profiles in a micro-structured copper plate taken with an infrared camera are in good agreement with theory. Cloaking is conceptually distinct from mere isolation. In optics (thermodynamics), an object can easily be isolated from its surrounding by wrapping a metal (isolator) around it. However, the flow of light (heat) will be influenced by the presence of the isolation. A true cloak not only isolates but also makes the isolation invisible.

The basic idea of transformation thermodynamics and transformation conduction is illustrated in Fig.~\ref{fig1_resistor-grid}. The resistors connecting adjacent Cartesian grid points in panel (a) can either be Ohmic or thermal. For metals, according to the Wiedemann-Franz law, the two types of resistors are simply proportional to each other at a given temperature. Upon performing the coordinate transformation given in \cite{Pendry2006} in panel (b), the grid lines become distorted and an empty region is opened in the middle. However, this distortion does not influence the flow of the electrical or the heat current outside the structure, i.e., for $r > R_2$, at all, provided that adjacent grid points are still connected by identical resistors (and assuming, as usual, that the connecting wires have vanishing resistance). Thus, the circular hole in the middle can be cloaked. To realize such a fictitious resistor distribution by an actual material distribution, it is interesting to note that the area density of resistors in panel (b) of Fig.~\ref{fig1_resistor-grid} is neither homogeneous nor isotropic. Close to the inner edge of the cloak at radius $r = R_1$  in panel (b), the density of parallel resistors providing current flow in the azimuthal direction is large; hence the electric or the thermal azimuthal conductivity is large. In contrast, only few parallel but more serial resistors in this region provide conduction in the radial direction. Hence, the radial conductivity is smaller and even vanishes towards the inner radius $r = R_1$. Mathematically \cite{Guenneau2012}, Guenneau and coworkers derived the following approximate form for the azimuthal component
\begin{equation}
\kappa_\theta = \kappa_0 \left( \frac{R_2}{R_2 - R_1} \right)^2 \geq \kappa_0
\label{eqn:kappa_theta}
\end{equation}
and the radial component
\begin{equation}
\kappa_r = \kappa_0 \left( \frac{R_2}{R_2 - R_1} \right)^2 \left( \frac{r-R_1}{r} \right)^2 \leq \kappa_0
\label{eqn:kappa_r}
\end{equation}
of the heat conductivity tensor $\overset{\leftrightarrow}{\kappa}$ in the radius interval $R_1 \leq r \leq R_2$. Here, $\kappa_0$ is the isotropic heat conductivity of the surrounding of the cloak. Furthermore, they found \cite{Guenneau2012} that the product of the effective mass density $\rho$ and the effective specific heat $c$ is approximately spatially constant, i.e., $\rho\,c = \text{const}$.

\begin{figure}
  \centering
  \includegraphics[scale=0.75]{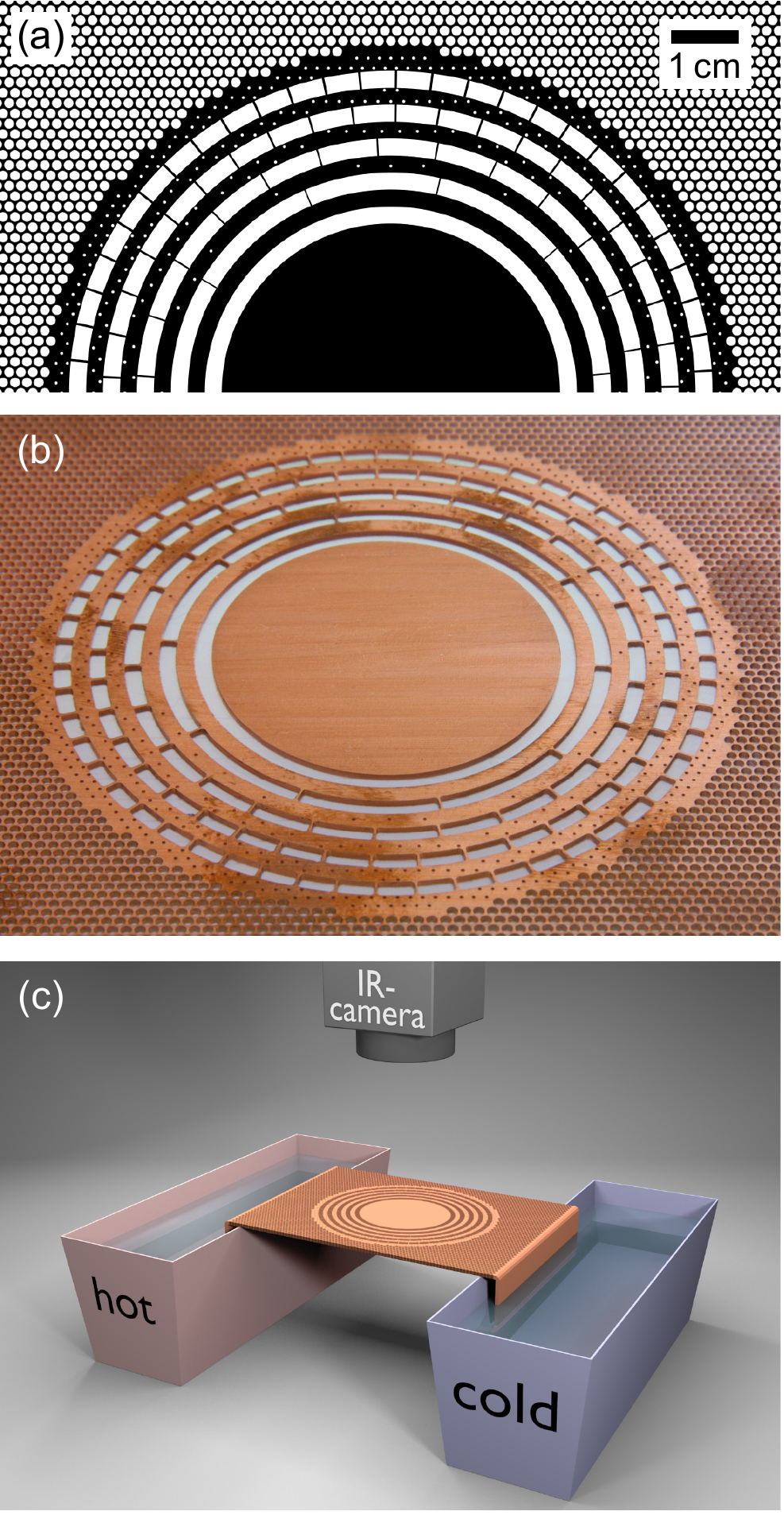}
  \caption{(a) Blueprint of the designed thermal cloak (only one half is shown due to symmetry). The black regions are bulk copper, the white regions polydimethylsiloxane (PDMS) with heat conductivities of $394\,\text{W}/(\text{Km})$ and $0.15\,\text{W}/(\text{Km})$, respectively. As illustrated in Fig.~\ref{fig1_resistor-grid}\,(b), the radial and azimuthal components of the effective local heat conductivity follow a recipe pre-described by the coordinate transformation. (b) Photograph of the fabricated thermal cloak and (c) scheme of the experiment, results of which are shown in Fig.~\ref{fig3_experimental_results}.}
  \label{fig2_structure_design}
\end{figure}

These material distributions still need to be mapped onto a practical metamaterial micro-structure exhibiting such pre-described anisotropic heat conduction. Inspired by our previous experiments on mechanics \cite{Stenger2012}, we choose 10 equally thick rings with alternating large and small effective heat conductivities with a contrast that gradually decreases from the inner radius $R_1$ to the outer radius $R_2$. The heat conductivities $\kappa_i$ of the rings ($i=$\ 1--10 from the inside to the outside of the cloak) are determined using formulae (24) given in Ref.~\cite{Guenneau2012} to obtain the best match to (\ref{eqn:kappa_theta}) and (\ref{eqn:kappa_r}) in the region with $R_1 \leq r \leq R_2$. In units of $\text{W}/(\text{Km})$, this leads to $\kappa_1 = 0.15$, $\kappa_2 = 394.0$, $\kappa_3 = 2.91$, $\kappa_4 = 390.0$, $\kappa_5 = 11.26$, $\kappa_6 = 382.7$, $\kappa_7 = 19.02$, $\kappa_8 = 375.0$, $\kappa_9 = 26.38$, and $\kappa_{10} = 367.6$. Intuitively, we exploit the fact that heat conduction is large along the direction of a ring with large heat conductivity, i.e., in the azimuthal direction, whereas heat conduction is smaller in the perpendicular radial direction. The effective anisotropy depends on the contrast within a pair of rings. To act as an effective material, the spacing between the rings needs to be small compared to the typical scale of temperature gradients, i.e., small compared to the thermal diffusion length.

To obtain these different heat conductivities, we design a composite structure made by drilling  holes into a copper plate and filling them with polydimethylsiloxane (PDMS). The corresponding (room-temperature) bulk heat conductivities are $\kappa_\text{Cu} = 394\,\text{W}/(\text{Km})$ and $\kappa_\text{PDMS} = 0.15\,\text{W}/(\text{Km})$. The latter is more than three orders of magnitude smaller than the former. The metal area fraction $f_i$ in ring number $i$ is obtained from the effective-medium formula $\kappa_i = f_i\,\kappa_\text{Cu} + (1-f_i)\,\kappa_\text{PDMS}$. The product $\rho\,c$, which ought to be constant, cannot be adjusted independently in our approach. The bulk values for copper and PDMS are $\rho\,c = 3.4\,\text{MJ}/(\text{Km}^3)$ and $\rho\,c = 1.4\,\text{MJ}/(\text{Km}^3)$. We find, however, that the local effective specific heat merely varies by about 30\,\% in the radius interval $[R_1,R_2]$ for our parameters. To compensate for this approximation, based on a numerical optimization, we choose the effective heat conductivity of the surrounding $\kappa_0 = 85\,\text{W}/(\text{Km})$.

The absolute size of all mentioned features is not relevant for the behavior of the thermodynamic cloak. If we replace $\vec{r} \rightarrow s\,\vec{r}$ and $t \rightarrow s^2\,t$ with a dimensionless factor $s$, the heat conduction equation remains unchanged. This means that, for example, reducing the size of all lateral features by a factor of 10 leads to a thermal time constant which is smaller by a factor of 100---of course provided all material properties remain unchanged. We choose $R_1=2.5\,\text{cm}$ and $R_2=5\,\text{cm}$, leading to a ring thickness of 2.5\,mm. The thickness of the plate is not important for the heat-conduction problem of interest, but the thickness does influence the strengths of artifacts. For large plate thicknesses, the plate's volume-to-surface ratio is large and, hence, heat conduction/convection to air is small. However, at some point the overall heat capacity of the plate is so large that it becomes difficult to define temperature ``baths'', which ought to have heat capacities much larger than that of the plate. We find that a plate thickness of 2\,mm provides a good trade-off. 

\begin{figure}
  \centering
  \includegraphics[scale=0.8]{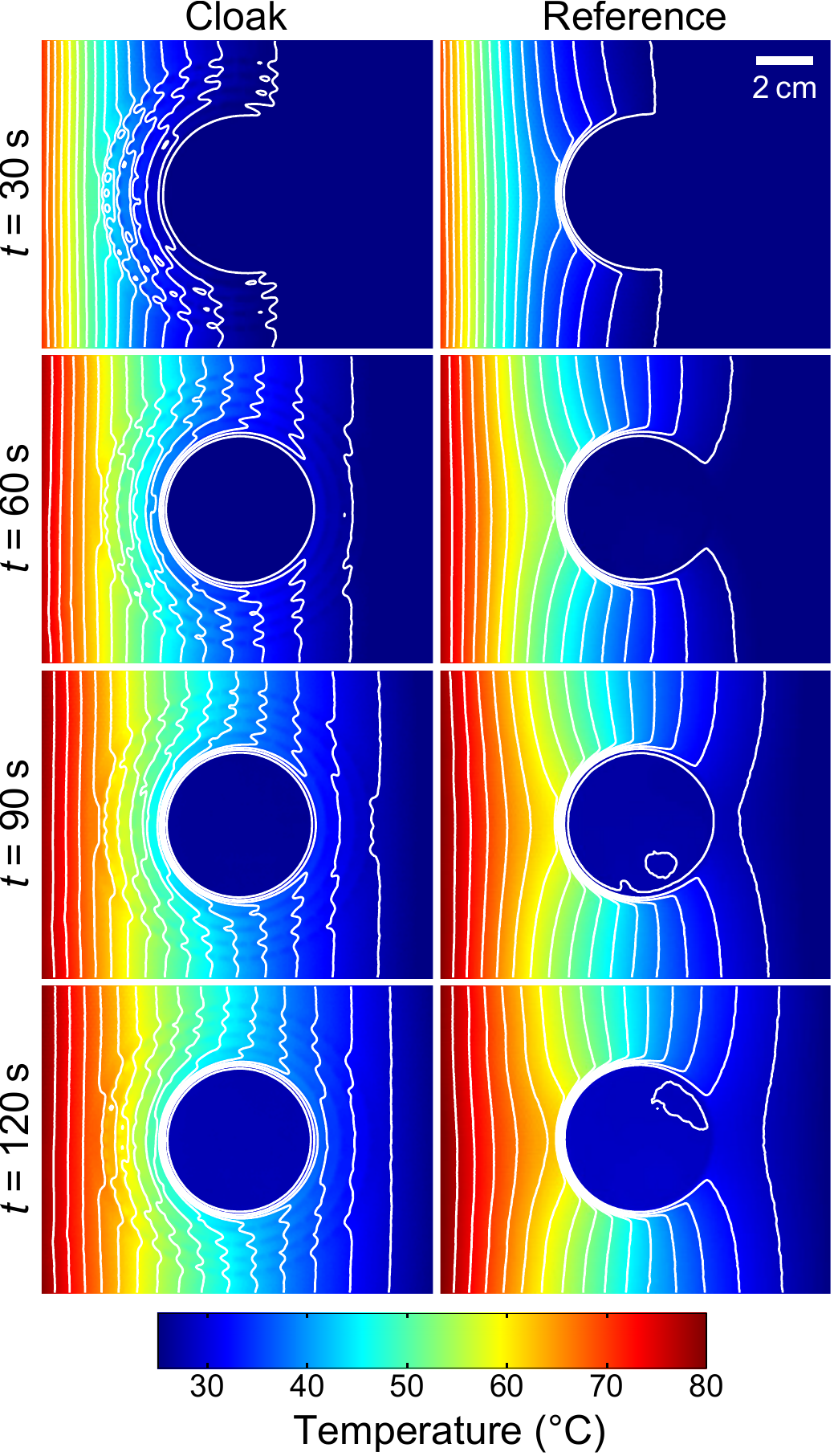}
  \caption{Measured temperature distributions at different times $t$ as indicated. At $t = 0\,\text{s}$, we start from a homogenous room-temperature profile. Results for the complete thermal cloak shown in Fig.~\ref{fig2_structure_design}\,(b) are shown in the left-hand side column, results for a simplified reference structure with only a thermally isolating ring around the central copper region are depicted in the right-hand side column. The white curves are iso-temperature lines (equidistant in steps of $3^\circ\,\text{C}$) corresponding to the temperature profiles depicted on a false-color scale.}
  \label{fig3_experimental_results}
\end{figure}

In the fabrication of the structure, we start with a bulk copper plate and machine the pattern shown in Fig.~\ref{fig2_structure_design}\,(a) into it. For some of the rings, hexagonal arrays of holes (with 0.5\,mm diameter) are drilled. For other rings, the very low metal filling fractions inhibit doing that. In these rings, we rather realize the low metal filling fraction by azimuthally evenly distributed thin radial bars. This step makes heat conduction in these individual rings anisotropic. However, our experiments show that this approximation has little if any detrimental effect on the performance of the system. To compensate for all approximations made, we adjust the metal filling fraction in the surrounding. Based on an experimental optimization, we choose an hexagonal lattice of holes with lattice constant 1.76\,mm and diameter 1.5\,mm, leading to $f_0 = 34.2\,\%$. Thereafter, the holes are filled with PDMS, which is polymerized afterwards. Furthermore, we have thermally isolated both surfaces of the composite plate by an approximately $100\,\upmu\text{m}$ thin layer of PDMS. As a result, heat conduction/convection by air is significantly reduced.

To measure the transient temperature profiles via Planck thermal emission using a conventional infrared heat camera (\textsc{Flir} A320), it is important to achieve nearly 100\,\% $\text{absorbance} = \text{emissivity}$ according to Kirchhoff's law. Fortuitously, in sharp contrast to the highly reflective copper itself, the $100\,\upmu\text{m}$ thin PDMS layer on the top of the plate is nearly ``black'' for the wavelengths seen by the thermal heat camera. Thus, we need no additional coating of the surface. The absolute temperatures in the temperature profiles shown below are derived from the camera data assuming 99\,\% emissivity (the standard value). A photograph of one of the fabricated structures is depicted in Fig.~\ref{fig2_structure_design}\,(b). Local heating on the left-hand side of the plate is achieved by a tank filled with hot water (see panel (c) in Fig.~\ref{fig2_structure_design}). The heat capacitance of the filled tank is about two orders of magnitude larger than that of the composite plate, qualifying it as a temperature bath in the thermodynamic sense. The right-hand side of the plate is immersed in an identical tank filled with room temperature water. The remaining two sides of the plate are left open (i.e., isolated). For waves, such termination would not be possible because waves are generally reflected at an open edge. 

Figure~\ref{fig3_experimental_results} shows results of our experiments for the complete cloak (left-hand side column) and for a simplified isolating structure (right-hand side column) at various times $t$ (different rows) after exposing the plate to the heat baths at $t = 0\,\text{s}$. Prior to that, the plate was kept under room temperature conditions for a sufficiently long time such that it exhibited a spatially constant temperature at $t = 0\,\text{s}$. The temperature in the plane of the plate is shown on a false-color scale. The cloak obviously successfully fulfills its task in that the central region is colder than its surrounding. Furthermore, the white iso-temperature lines on the downstream side of the cloak in Fig.~\ref{fig3_experimental_results} are nearly vertical, indicating a temperature distribution as if nothing was there. However, one must be cautious because part of the desired and achieved effect can also be obtained by a simple thermal isolation around the circular central region. Thus, we have fabricated a corresponding reference sample with a solid copper region and the first PDMS ring around it (parameters like for the complete cloak). The surrounding is a homogeneous perforation just like for the cloak, i.e., this reference structure has no further rings. The result depicted in the right-hand side column of Fig.~\ref{fig3_experimental_results} also shows a central region that is cooler than its surrounding. This finding is very much different from light waves or mechanical waves due to the fundamental difference between hyperbolic and parabolic differential equations emphasized in the introduction. The optical counterpart of the weakly heat conducting PDMS ring would be an opaque metal film. In mechanics, the analog would be a rigid wall. As a result, one gets scattering, shadowing, and wave-front distortions. These effects have, e.g., been observed in Ref.~\cite{Stenger2012} for a comparable two-dimensional scenery. In contrast, in Fig.~\ref{fig3_experimental_results}, all of these effects are absent. One merely finds distortions of the iso-temperature curves, which may be seen as the counterpart of wave fronts. The iso-temperature curves in the homogeneously perforated region near the center of the reference structure are curved towards the center. Here, the heat flux is simply perpendicular to the iso-curves. In sharp contrast, the iso-temperature curves for the complete cloak in Fig.~\ref{fig3_experimental_results} exhibit the opposite curvature, namely away from the cloak, both on its left and on its immediate right. The local heat flux is still normal to the iso-temperature curves. Averaged over the fine wiggles due to the metamaterial features, however, the heat flux results from the product of the local anisotropic heat conductivity tensor and the local temperature gradient. 

\begin{figure}
  \centering
  \includegraphics[scale=0.8]{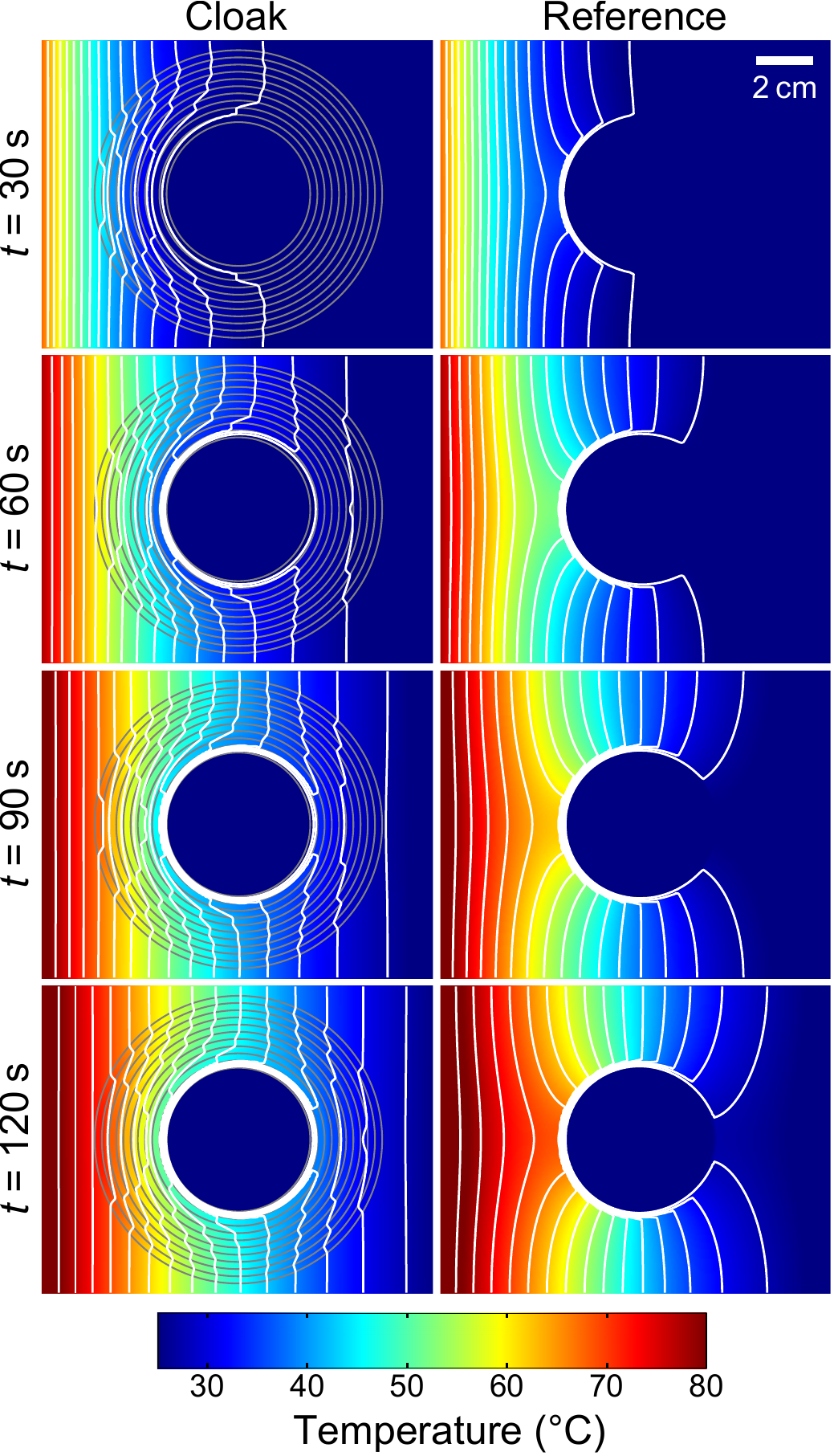}
  \caption{Calculated temperature distributions shown as the experimental results in Fig.~\ref{fig3_experimental_results}. For the cloak in the left-hand side column, the ring structure is indicated for clarity.}
  \label{fig4_numerical_results}
\end{figure}

Generally, our experiments in Fig.~\ref{fig3_experimental_results} agree well with the corresponding numerical calculations (\textsc{Comsol}) shown in Fig.~\ref{fig4_numerical_results}. The heat baths are mimicked by fixed temperatures. Furthermore, the calculations assume isotropic homogeneous rings with the prescribed heat conductivities given above and do not account for heat conduction to the surrounding air. The latter aspect likely explains the remaining discrepancies. We refrain from showing the modeling accounting for this aspect because it is not part of the cloak design. Applications would likely use smaller structures than our model anyway. Hence, as discussed above, the time scales become shorter and heat transport via air (especially via convection) would play an even smaller role. Clearly, by design, the cloak recovers the overall downstream heat flow of the homogenously perforated plate under transient as well as under static conditions (see Supplemental Material \cite{SUPP}). However, in the static or long-time limit, due to the finite heat conductivity of the innermost isolating ring, the to-be-protected inner region does eventually heat up \cite{SUPP}. Hence, thermal protection works only transiently.

In conclusion, using the theoretical concepts of transformation thermodynamics, we have realized a cloak that molds the dynamic flow of heat around an object as if no object was there. Such thermal cloaks might find, e.g., applications in temporarily protecting sensitive regions in an electrical circuit or chip from excessive heating. More importantly, our experiments also demonstrate that the ideas of transformation optics go well beyond systems exhibiting waves and also work for transient heat conduction.

We thank Frank Landh\"{a}u\ss er, Mario Nusche, and Johann Westhauser (KIT) for help in the fabrication, Christoph Kaiser (KIT) for lending to us the infrared heat camera, and Jonathan
M\"{u}ller (KIT) for stimulating discussions regarding the cloak design. We
acknowledge support by subproject A1.5 of the DFG-Center for Functional
Nanostructures (CFN).

%

\end{document}